\documentclass[aps,prl,twocolumn,groupedaddress,showpacs,draft]{revtex4}
\begin{document}

\title{An Isotropic Metric.}
\author{Joseph~D.~Rudmin}
\email{rudminjd@jmu.edu}
\affiliation{James Madison University}
\address{JMU ISAT, MSC4106, 701 Carrier Dr., Harrisonburg, VA 22807}
\date{May 29, 2007} 

\begin{abstract}
An isotropic metric for a black hole and a better vacuum condition 
\mbox{$\nabla^{2} V_{G}=0$} are presented which yield distinct terms 
for the energy densities of ordinary matter and gravitational 
fields in the Einstein tensor 
\mbox{($G^{44}=-g^{2}(2\nabla^{2} V_{G} +(\nabla V_{G})^{2} )$).} 
This model resolves an inconsistency between electromagnetism and gravity 
in the calculation of field energy.  Resolution of this inconsistency 
suggests a slight modification of the Einstein equation to 
\mbox{$gG^{\mu \nu} = 8\pi G T^{\mu \nu}$}.
\end{abstract}

\pacs{04.20.Cv, 04.70.-s, 04.20.Jb, 04.20.-q}
\keywords{gravitation theory, relativity, Einstein equation, Einstein tensor,
black holes, gravitational distortion, geodesics, negative energy condition,
massless fields, negative mass, Schwarzschild metric, isotropic metric,
electromagnetic fields}

\maketitle
\section{Introduction.}
One can calculate the energy in the electric field of a shell of charge two 
different ways, which give the same result:  One can integrate the square of 
the electric field over all space around the shell of charge, or one can 
integrate the work done in moving the charges to the shell.  
For electromagnetism, like charges repel.  So one does positive work to assemble 
a sphere of charge, like compressing a spring.  Thus an electric field has 
positive energy and positive mass.  For gravity, like charges 
attract.  So one does negative work to assemble a spherical shell of mass.  
Thus, gravitational fields should have negative energy and negative mass.  
The vector or tensor nature of the fields has no signficance in this
calculation since the difference in behavior is negligible at least in the
weak field limit.  The energy stored is force through distance.

Analogies between gravitational and electromagnetic fields are usually
explored in the linear approximation for gravitational fields.  (See, for
example, Chapter 3 of the text by Ohanian and Ruffini. ~\cite{Ohanian1} )
Such expositions may recognize that in the linear approximation, the
laplacian of the field should be zero in the absence of matter, and that
a gravitational field should contribute a term to the energy density which has
a form of the square of the gradient of the gravitional potential.  
(e.g. ff p.147-148.)  However, when the full equations are developed,
the curvature tensor $R^{\mu \nu}$ is assumed to be zero in the absence 
of matter.  As a result $G^{\mu \nu}$ is also zero, implying that gravitational
fields have no energy density.  This assumption goes back a long way.

Almost all papers on gravitational fields over the last ninety years assume 
that mass density is always nonnegative.  For example, the Einstein
tensor for the Schwarzschild metric is zero, making its fields massless.  
The only papers that have admitted the concept of negative mass
(e.g. Visser et. al.~\cite{MVisser1, DHochberg1} and Kip Thorne et. al
~\cite{Thorne1})
have done so as a purely theoretical tool to explore the concept 
of wormholes, although weaknesses in this assumption have been identified 
~\cite{Barcelo1}.  This assumption of nonnegative mass is the energy 
conditions used for all metrics in common use.  Therefore, those metrics may 
be incorrect.

An isotropic metric for a black hole is presented here for which $G^{\mu \nu}$
has distinct terms for ordinary matter and gravitational fields.  This 
$G^{\mu \nu}$ is derived in the usual way from the metric tensor $g_{\mu \nu}$.
When $g_{\mu \nu}$ is isotropic, one can define a gravitational potential, and 
then $G^{\mu \nu}$ derived from it takes the form of a difference between 
the laplacian and the square of the gradient of the potential.
The Einstein Tensor is a purely geometric quantity.  Instead of attributing
the entire $G^{\mu \nu}$ to ordinary matter, only the laplacian of the
gravitational potential is attributed to ordinary matter.  The remainder,
which has the form of an energy density of a field is attributed to the
contribution of gravitational fields.  It should not be surprising that each
term of $G^{\mu \nu}$ transforms properly under a lorentz transformation, 
because it is incidental whether each term is locally zero.  An isotropic 
metric has the additional feature that world lines do not cross event horizons,
thus avoiding interactions with regions where physical models break down.

\section{Metric and Einstein Tensor.\label{Metric1}}
For an isotropic metric with gravitational distortion $g$, in the rest frame 
of the source of the gravitational field,
\begin{equation} ds^{2} = -(g_{r})^{2}(dx^{2} + dy^{2} + dz^{2}) + dt^{2}/(g_{t})^{2}. 
\label{eqds2} \end{equation}
Justification will be shown later for $g_{r}=g_{t}=g$.  
This metric differs from simply isotropic coordinates in that a sphere of
constant $r$ has a surface area of $4\pi g^{2}r^{2}$ instead of $4\pi r^{2}$.
Because the speed of light slows by a factor of $g^{2}$, the metric is {\em not} 
conformally flat, as will be shown later with the geodesics (Eq.~\ref{eq:geod1}).  

The gravitational distortion affects momentum and energy as well as distance and 
time.  For example, 
one could use a photon with a frequency matched to that of a clock 
in a gravitational well to carry information about the clock out of the well.  
Then, the gravitational distortion should affect energy the same as frequency.
As a photon of energy $M_{\gamma}$ climbs out of the gravitational potential, 
$g \cdot d(M_{\gamma}V_{G}) = -d(gM_{\gamma})$, which yields $g=\exp{(-V_{G})}$.  
Then the Einstein tensor, in spherical coordinates, derived from the isotropic 
metric of the length differential (Eq.~\ref{eqds2}) is:
\begin{equation} \begin{array}{rl} G^{\mu \nu} = & \hspace{-0.1in}
 \left( \begin{array}{cccc}
(\nabla V_{G}/g)^{2} & \hspace{-0.2in} 0 & \hspace{-0.3in} 0 & 0 \\
0 & \hspace{-0.2in} -(\nabla V_{G}/rg)^{2} & \hspace{-0.3in} 0 & 0 \\
0 & \hspace{-0.2in} 0 & \hspace{-0.3in} -(\nabla V_{G}/rg\sin{\theta})^{2} & 0 \\
0 & \hspace{-0.2in} 0 & \hspace{-0.3in} 0 & G^{44} \end{array} \right) \end{array} \end{equation}
where $G^{44}=-g^{2}(2\nabla^{2} V_{G} +(\nabla V_{G})^{2} )$.  
Einstein's equation is $G^{44}=8\pi G T^{44}$, where $T^{44}=\rho$ is the 
mass density.  Because \mbox{$V_{G}<0$}, \mbox{$-g^{2}\nabla^{2} V_{G}>0$}.  
This term represents mass density of ordinary matter.  The term 
\mbox{$-g^{2}(\nabla V_{G})^{2} < 0$} represents the mass density of 
gravitational fields.  Both the gradient and laplacian are calculated 
with the metric scaling the coordinates in the usual manner.

If one admits azimuthal as well as radial variation for $g$, then $G^{44}$ 
still retains the same form.  When expanded,
\[ G^{44}= \left(
2\frac{g_{,rr}}{g}+\frac{4g_{,r}}{rg}-\left( \frac{g_{,r}}{g} \right) ^{2}
\right) + \]
\begin{equation} + \frac{1}{r^{2}} \left(
2\frac{g_{,\theta \theta}}{g}+2\frac{g_{,\theta}}{g}\cot{\theta}
-\left( \frac{g_{,\theta}}{g} \right) ^{2} \right). \end{equation}
A comma indicates partial derivatives with respect to the coordinates that 
follow it.  The squared terms are the square of the gradient of the potential.
All other terms are the laplacian of the potential.

\section{Mass Reconciliation.\label{MassRec}}
The negative mass of the gravitational fields inferred by analogy with 
electromagnetic fields should be quantified.  Suppose one attempts such a 
calculation, to determine the form for the gravitational distortion $g$ in 
this model.  In a vacuum, $\nabla^{2} V_{G}=0$, which is 
\begin{equation} 2\frac{g_{,rr}}{g}+\frac{4g_{,r}}{rg}=0. \end{equation}
With $g=1$ and $V_{G}=-GM/r$ for large $r$,
the solution to this equation is 
\begin{equation} g=1+\frac{GM}{r}, \end{equation}
where $M$ is the mass of the shell of matter.  
The remaining term in $G^{44}$ is the mass of the gravitational fields.  
With spherical symmetry, 
\begin{equation} G^{44}=-g^{2} \left( \nabla V_{G} \right)^{2} 
= -\left(\frac{g_{,r}}{g}\right)^{2}.  \end{equation}
Total mass of the gravitational fields
\begin{equation}M_{G}=2\int^{\infty}_{r=r_{0}} \frac{G^{44}4\pi r^{2}dr}{8\pi G}
=\frac{1}{G} \int^{\infty}_{r=r_{0}} G^{44} r^{2} dr. \end{equation}
If the energy of the gravitational fields is that which is released in 
assembling the shell of mass $M$, then $M_{G}=MV_{G}=-M \ln{g}$, in analogy
with assembling a shell of electric charge.  When 
one equates these two calculations of the mass, 
\begin{equation} -GM \ln{g} = \int^{\infty}_{r=r_{0}} 
-\left( \frac{g_{,r}}{g} \right) ^{2} r^{2} dr. \end{equation}
\begin{equation} \frac{-GMdr}{r^{2}} = \frac{dg}{g}. 
\hspace{0.3in} \frac{GM}{r} = \ln{g} = -V_{G}. \end{equation}
This result does not satisfy the vacuum condition $\nabla^{2} V_{G}=0$.  
Apparently, the problem is over defined.  

To solve this paradox, one might allow the distortion for the time coordinate, 
$g_{t}$, to differ from that for the space coordinate, $g_{r}$, and apply the 
vacuum condition only to $g_{r}$.  With this substitution, $G^{44}$ retains the 
desired form, $G^{44}=-g_{t}^{2}(2\nabla^{2} V_{G} +(\nabla V_{G})^{2} )$, and 
$g_{t}$ appears only as a scaling factor, and not in the operators on $V_{G}$.  
If one admits further anisotropy, then $\nabla^{2} V_{G}$ and $(\nabla V_{G})^{2}$ 
no longer appear as distinct terms in any components of $G^{\mu \nu}$.
As shown above, $g_{r}=1+GM/r$ to satisfy the vacuum condition.  Then 
equating the two ways of calculating mass results in $g_{t}^{2}=g_{r}^{3}$.  
It is reasonable to conclude that instead of $G^{\mu \nu}=8\pi G T^{\mu \nu}$, 
the Einstein Equation should be 
\begin{equation} gG^{\mu \nu}=8\pi G T^{\mu \nu}.\end{equation}  
The left side of this equation retains its tensor properties because $g$ is 
a scalar.  As a result, three factors of $g$ appear on the left side of the
Einstein equation, multiplying the laplacian and squared gradient when 
$G^{\mu \nu}$ is expanded.  Qualitatively, one can understand these three 
factors of $g$ as scaling the volume differential of the energy density.  
A factor of $1/g$ should scale the energy, but it is cancelled by an additional 
factor of $g$ which scales the gravitational potential.
As a bonus, because mass reconciliation then yields 
$g_{t}=g_{r}$, both $g_{t}$ and $g_{r}$ satisfy the vacuum condition.
For the rest of this paper, $g_{t}=g_{r}=g$.

\section{Geometry Near a Black Hole.\label{BHGeom}}
As one descends into this black hole with isotropic gravitational distortion 
$g=1+GM/r$, it becomes increasingly self similar, since both $M$ and $r$ scale 
the same way with the gravitational distortion.  Locally, the circumference 
asymptotically approaches $2\pi GM$, and the remaining distance to the event 
horizon asymptotically approaches $GM$.

To calculate the geodesics, one integrates the local time for a photon 
to travel between two points, factors out the constant $g^{2} c$, and
applies the Euler-Lagrange equations to the integrand.
\newcommand{\dssq}{\left( \left( r_{,\theta} \right) ^{2}+r^{2} \right) }
\newcommand{\dslen}{\sqrt{ \dssq }}
\begin{equation} \Delta T = \int \frac{dt}{g} = \frac{1}{g^2 c} \int g 
\dslen \hspace{0.1in} d\theta. \end{equation}
The resulting geodesics are:
\begin{equation} 0=r^{2} r_{,\theta \theta}-2r \left( r_{,\theta} \right) ^{2}
-r^{3}-\frac{g_{,r}}{g} \dssq ^{2}. \end{equation}
For $g=1+GM/r$,
\begin{equation} r_{,\theta \theta}=\frac{2\left( r_{,\theta} \right) ^{2}}{r}+r
-\frac{GM}{r(r+GM)}\dssq ^{2} {\label{eq:geod1}} \end{equation}
The right most term distinguishes these geodesics from those for flat space.  
It deflects the path of light toward the gravitational potential.  
The time for a photon to reach the event horizon at the center is infinite 
whether measured by an outside observer:  
\begin{equation} \Delta T=\frac{1}{g^{2} c}\int_{r_{1}}^{r_{2}} g ds \approx
\frac{GM}{g^{2} c}\log{\frac{r_{2}}{r_{1}}} \label{eq:time1} \end{equation}
or in a frame descending into the gravitational well: 
\begin{equation} \Delta T=\frac{1}{g^{2}c}\int_{r_{1}}^{r_{2}} g^{2} ds \approx
\frac{GM}{g^{2} c} \left( \frac{1}{r_{2}} - \frac{1}{r_{1}} \right) 
\label{eq:time2} \end{equation}

\section{A Massless Metric.\label{0MasMet}}
For purposes of comparison, the following gravitational distortion describes a 
distribution of matter contrived to exactly cancel the negative mass of 
gravitational fields everywhere outside the event horizon:
\begin{equation} g=\frac{r}{r-GM}=\frac{1}{1-\frac{GM}{r}}. \end{equation}
Although $G^{44}=0$ for this metric, $G^{11}$, $G^{22}$, and $G^{33}$ 
are all nonzero.  So, $G^{\mu \nu}$ is not the same as for flat space.  

An event horizon resides at $r=GM$.  Substitution of $g$ into the equations 
for time of travel (Eq.~\ref{eq:time1}, \ref{eq:time2}) shows that objects still 
do not cross the event horizon.  At a radius $2GM$, this metric has a waist, 
where the circumference is a minimum.  
Circumference \mbox{$l_{C}=2\pi rg=2\pi r^{2}/(r-GM)$} which has a minimum 
value of $8\pi GM$.  The total distance between two points at different depths
\[ s=\int \frac{rdr}{r-GM} 
= \int_{r_{1}}^{r_{2}} \left( 1+\frac{GM}{r-GM} \right) dr = \]
\begin{equation} 
= (r_{2}-r_{1}) + \log \left( \frac{r_{2}-GM}{r_{1}-GM} \right). \end{equation}
Putting $r_{1}=2GM$ at the waist and $r_{2}$ inside the waist shows that 
the circumference grows exponentially with depth:  
\begin{equation} s\approx \log \left( \frac{r_{2}}{GM}-1 \right). \end{equation}
\begin{equation} l_{C} =\frac{2\pi r_{2}^{2}}{r_{2}-GM} 
\approx \frac{2\pi GMe^{2s}}{e^{s}-1} \approx 2\pi GMe^{s}. \end{equation}

Thus, the distribution of mass makes the space more expansive there than it 
would be if the matter were absent.  One might interpret matter as an excess 
of volume within a surface area.

\section{Conclusions and Implications.}
Not only does an isotropic metric result in gravitational fields with negative 
mass, as one should expect, it offers a number of other advantages over the 
Schwarzschild metric.  As shown above, an isotropic metric results in a very 
symmetric form for the Einstein tensor, with distinct terms for ordinary mass 
and gravitational fields.  Objects do not cross event horizons.  
A large amount of free energy available to objects falling in the halo of 
a black hole might nucleate cosmoses.  For example, the massless metric just 
shown illustrates how the presence of an energy density induces expansion.  
Since this metric is isotropic, it can accommodate the 
nucleation of isotropically expanding cosmoses in the halo of a black hole in a 
way that the Schwarzschild metric cannot.  An isotropic metric also terminates 
electromagnetic field lines in a way that the Schwarzschild metric and Kerr 
metrics cannot:  Deep enough into the halo of the black hole, the circumference 
and surface area increase, thus causing electromagnetic field strengths decrease.  
Projected out, it appears that a charge density resides in the halo of a charged 
black hole.  The termination of field lines provides cutoffs for fields, 
limits field energies, and might accommodate 
general relativistic models for the masses of the electron, muon, and tauon.



\begin{thebibliography}{99}
\bibitem{Barcelo1} C. Barcelo and M. Visser, Int. J. Mod. Phys D. {\bf 11} (2002) 1553.
\bibitem{MVisser1} C. Catto\"{e}n and M. Visser, Class. Quant. Grav. {\bf 22} (2005) 4913.
\bibitem{DHochberg1} D. Hochberg and M. Visser, Phys. Rev. Lett. {\bf 81} (1998) 746.
\bibitem{Thorne1} M.S. Morris and K.S. Thorne, Am. J. Phys. {\bf 56} (1988) 395.
\bibitem{Ohanian1} H. Ohanian and R. Ruffini, Gravitation and Spacetime, 2nd ed. (1994).
\end{thebibliography}
\end{document}